\documentclass[toc]{proar}

\title{Properties of Confinement in Holography}

\ShortTitle{Confinement in Holography}

\author{
Dimitrios Giataganas
\\
 Department of Nuclear and Particle Physics,
Faculty of Physics, University of Athens,
Athens 15784, Greece
\\
        E-mail: \email{dgiataganas@phys.uoa.gr}}

\abstract{We review certain properties of confinement with added focus on the ones we study with holography. Then we discuss observables whose unique behavior can indicate the presence of confinement. Using mainly the Wilson loop in the gauge/gravity formalism, we study two main features of the QCD string: the string tension dependence on the temperature while in the confining phase, and the logarithmic broadening of the flux tube between the heavy static charges that turns out to be a generic property of all confining theories. Finally, we review the $k$-string bound state and we show that for a wide class of generic theories the $k$-string observables can be expressed in terms of the single meson bound state observables.}

\begin{document}

\section{Introduction to Aspects of Confinement}

\subsection{Confinement}

Confinement is a property of QCD which implies a growing force between color charged particles when increasing their separation distance. The gluonic mediators of the strong force in the confined phase between a quark bound state are constrained in a tube like region. The basic characteristic of confinement is that the potential energy between a static heavy quark-antiquark potential is a linearly increasing function with the separation between the quarks. The proportionality constant is defined as the string tension and its value indicates how strong is the force. The linear static potential corresponds to an area law for Wilson loop expectation values.

The linear growth in the potential is not indefinite. When the separation between the quark bound state passes a critical value then the confining tube breaks and a creation of two new bound states involving lighter quarks is energetically preferable. The potential of the initial bound state at this phase becomes flat. In practice this phenomenon implies that in the confining phase the color charged particle states are absent, a statement that can also serve as an alternative definition of confinement.

\subsection{The Property of the QCD Spectrum}

The hadronic bound states of baryons and mesons of the same flavor quantum numbers, have spin that depends linearly with their squared mass.  This is a striking feature and can be approximated by the spinning stick model, while rotating around its midpoint. The simple exercise gives for the angular momentum and the mass the relation
\begin{equation}
J=\frac{m^2}{2 \pi \sigma}~,
\end{equation}
with the constant $1/\prt{2 \pi\sigma}$ being the Regge slope. Comparing and fitting the result to the data, we obtain an approximate value of the string tension $\sigma\simeq 0.9 GeV/fm$. Notice that the mapping of the Regge trajectories to the solid stick problem is only a naive approximation. However, it is enough to provide the physical intuitive picture that the QCD interactions have essential differences compared to other interactions and approach an effective description  by the "QCD string" and the formation of the color electric flux tube.

\subsection{How to Probe the Confinement}

To examine whether or not the system is in a confined phase we should interact somehow with it and make a measurement. There are certain observables that can be used for this purpose, where few of them we list below.

\subsubsection{The Wilson Loop and the Static Potential}

The expectation values of the Wilson loop observable is possibly the most commonly used quantity to probe confinement. The idea is to introduce in the theory a static heavy meson probe, consisting of a static quark and antiquark separated by a distance $L$. The two fermions are considered to be very heavy, therefore as the time passes the quarks remain fixed in space, forming two infinite length -in good approximation- lines, as trajectories in the space-time.

The Wilson loop is defined as the trace in the given representation of the path ordered exponential
 \be
W\prt{C}\propto Tr P e^{i g \oint_C dx^\m A_\mu\prt{x}}~,
\ee
where the integral is taken along the orthogonal loop $C$ and $A_\mu$  are the gauge fields of the theory. It can be shown that the Wilson loop expectation value can be expressed in terms of the energy eigenstates of the corresponding Hamiltonian. At the limit of large time $T$ the energy that dominates the sum is the interaction energy between the color sources
\be
\vev{W\prt{C}}=\sum_n|c_n|^2 e^{\Delta E_n T}\sim e^{-\Delta E_n T}\sim  e^{-V\prt{L} T}~, T\rightarrow \infty~.
\ee
Therefore, in the confined phase of the theory, $V\prt{L}\sim \s L$, implies an area law for the expectation value of the Wilson loops
\be
W\prt{C}\sim e^{-\s L T}=e^{-\s Area\prt{C}}~.
\ee
Notice that in certain large $N$ Wilson loop  computations may be a problem in the order of taking the large $T$ and $N$ limits, where special care need to be taken.

\subsubsection{Center Symmetry and Other observables}

There are several other probes of confinement additional to the Wilson loop that are related to the center symmetry. The center of the group is made by the group elements which commute with all the elements of the group. In the case of the $SU\prt{N}$ group, this is the $Z_N$ subgroup, $\{exp\prt{2 \pi in/N} I_N\}:=\{z_n I_N\}$, of $SU\prt{N}$. There are only $N$ representations of $Z_N$, and the N-ality $k$ of a given representation is given by the number of boxes $mod N$ in the appropriate Young Tableau. Having a matrix representation $M$ of $N$-ality $k$ of the Lie group, then the $zg$ element in the representation follows: $M[zg]=z^k M[g]$, where $z\in Z_N$, $g\in SU\prt{N}$. Notice that the $N$-ality, appear in the $k$-strings and their dual description of the higher representation Wilson loops.

A probe of confinement is the Polyakov loop, which is a Wilson loop winding around a lattice in the periodic time direction. This is an observable that is affected under center symmetry transformations unlike the action which remains invariant, and therefore its expectation value signals the center symmetry breaking. In an $SU\prt{N}$ gauge theory with $N$-ality zero matter fields, it is known that the unbroken $Z_N$ center symmetry phase corresponds to the zero Polyakov loop expectation value and can be identified with the confined phase. In higher temperatures the center symmetry is broken and the expectation values of the Polyakov loop is non-zero. This is the deconfined phase of the theory. Therefore the Polyakov loop may serve as an excellent observable of confinement.

There exist similar observables generalizing the idea of the Polyakov loop. This is done by generalizing the center symmetry transformation to a singular gauge transformation, which is not a true gauge transformation
\bea
U_0\prt{x,t}\rightarrow g\prt{x,t} U_0\prt{x,t}g\prt{x+1,t}~,\\
A_\mu\rightarrow g\prt{x}A_\mu g^\dagger\prt{x}-\frac{i}{\mbox{g}}g\prt{x}\partial_\mu g^\dagger\prt{x}~,
\eea
where $g\prt{x,t}$ is a non-periodic function in the periodic time direction, up to a center symmetry transformation $g\prt{x,T_0+1}=z^* g\prt{x,1}$. The singular gauge transformation does not leave the action invariant and create a singular loop of quantized magnetic flux, a center vortex.  To create the center vortex along the loop $C$ at time $t$ we need an operator $B\prt{C}$, the 't Hooft loop operator. To examine the behavior of the loop in the confining theory, we consider in 3-dim timeslice a loop $C$ that is topologically linked to another loop $C'$ (in a similar way two rings linked in a chain). The transformation then gives $B\prt{C}U\prt{C'}=z U\prt{C'} B\prt{C}~,$ with $ z\in Z_N$. $U\prt{C}$ is the Wilson loop holonomy defined as
\be
U\prt{C}\propto e^{i\oint_C dx^\mu A_\mu\prt{x}}~.
\ee
't Hooft \cite{tHooft:1977hy} showed that an area law behavior for $W\prt{C}$ implies a perimeter law fall of for $B\prt{C}$ which signals confinement.

There are additional probes of confinement which we do not mention in this short review, for a more complete discussion see \cite{Greensite:2003bk}. In this section we have shortly introduced the Wilson loops, where an area law implies confinement. The Polyakov lines, where a vanishing expectation value shows confinement. The 't Hooft loops where a perimeter law indicates that we are in the confined phase of the theory. The probes we have mentioned are different but can be thought as related to each other, along the lines we have briefly described.

\subsection{The String Tension}

The static potential between a heavy meson in zero temperature is a linearly increasing function at large separation distances $L$ with the proportionality constant given by the string tension $\sigma$ as
\be
V\prt{L}=\sigma L +\frac{c}{L}~.
\ee
For the linear behavior is responsible the shape of the world-sheet of the string, while the quantum fluctuations of the string generate the universal subleading L\"{u}sher term. At zero temperature the string tension is constant. Departing from the zero temperature and remaining in the confined phase of the theory, the string tension is decreased. More particularly for large separation lengths, the low temperature expansion of the string tension was found to be \cite{Pisarski:1982cn,deForcrand:1984cz}
\be
\s\prt{T}=\s_0-\frac{\prt{d-2}\pi}{6}T^2+\sum_{n\ge 3} \a_n T^n~,
\ee
where $\sigma_0$ is string tension in zero temperature. The second term is similar to the L\"{u}sher term of the static potential and is independent of the gauge group of the theory considered. Certain higher order terms can be found to be absent using a version of open-closed string duality \cite{Luscher:2004ib}.

\subsection{The Chromo-Electric Flux Tube}

The chromo-electric flux tube formed by the color electric field energy density between the static Q\={Q} pair is constrained in a tube-like region. Equivalently the tube can be thought as being created by the fluctuations of the color fields, which are constrained in tube-like shape.

The chromo-electric flux tube has interesting properties and is of almost equal importance with the string tension in confining theories. Its broadening  related to the width of the flux tube at the middle of the QCD string has been found to be increased logarithmically with the size of the bound state \cite{Luscher:1980iy}. This is a characteristic of the confining force and it could serve as a property of confinement itself. This phase is called the rough phase where the mean squared width of the flux tube diverges logarithmically with the inter-quark distance when it goes to infinity. Lattice computations have verified the logarithmic broadening using heavy mesons as probes in zero temperature, for example in \cite{Caselle:1995fh,Gliozzi:2010zv}.  Interestingly, it appears that the logarithmic broadening is a property of even the baryonic flux tube junction \cite{Pfeuffer:2008mz}. This clearly signals a universal behavior in chromo-electric field energy density in the confining phase, which is irrelevant of the probe used.
Departing from the zero temperature studies to the finite temperature, it has been found that the broadening changes to linear close to the phase transition \cite{Allais:2008bk,Gliozzi:2010zv}.

To compute the logarithmic broadening we need to study the effective width of the flux tube. This can be done by placing the heavy Q\={Q} bound state in our confining theory, which is associated with the  orthogonal Wilson loop $W\prt{C}$. The loop has one large time direction and another spatial with length $L$, being the size of the bound state. Along this direction is where the flux tube is formed. To probe the flux tube, we use another smaller Wilson loop $P\prt{c}$ on the plane above the Q\={Q} Wilson loop in the center of the large loop and we measure the flux passing through it. The mean square width of the flux tube is defined as
\be\label{wdefen11}
w^2=\frac{\int d x_\perp x_\perp^2 \cE\prt{x}}{\int d x_\perp \cE\prt{x}}\, ,
\ee
where $x_\perp$ are the transverse directions and $\cE\prt{x}$ is the chromo-electric field energy density
\be\label{wdefen22}
\cE\prt{x}=\frac{\vev{W\prt{C} P\prt{c}}-\vev{W\prt{C}}\vev{P\prt{c}}}{\vev{W\prt{C}}}~
\ee
measured with the use of the two Wilson loops $W\prt{C}$ and $P\prt{c}$.

\subsection{More Involved Bound States}

The meson bound states is a common way to probe confinement. More involved states in an $SU\prt{N}$ theory exist and have very interesting properties. Such states are the  $k$-strings, combinations of Q\={Q} pairs where each quark it is separated by the anti-quark by distance $L$ and each pair is separated by the other by distance $\e$, with $L\gg \e$. There is a common flux tube for the bound state with a static potential
\be
V=\sigma_k L+\frac{c}{L}~,
\ee
where $\sigma_k$ is the tension of the tube. The $k$-strings probe in a way also the interaction of the gluonic strings to each other and their study is expected to give an improved understanding of the dynamics of confinement. A direct question to ask, is if and how the string tension $\sigma_k$ is related to the fundamental string tension $\sigma_1$. The answer is still under debate, and there are at least two main proposals. The Casimir scaling \cite{Ambjorn:1984mb,casimir1}, where the expansion in large $N$ leads to the string tension in terms of $1/N$
\be
\sigma_k=k\prt{1-\frac{k-1}{N}+ \cO\left(\frac{1}{N^2}\right) }\sigma_1~,
\ee
and the sine formula, where the large $N$ expansion is in terms of $1/N^2$ powers
\be
\s_k= k\prt{1-\frac{\pi^2 \prt{k^2-1}}{6 N^2}+\cO\left(\frac{1}{N^4}\right)}\sigma_1~.
\ee
There are several models that find the Casimir or sine formula, for example \cite{cas2,cas3,cas4,cas5,cas6,cas7} and \cite{sin1,sin2}. However, the answer may be that the exact function of the string tension is none of these two proposed, although several other formulas have been excluded \cite{Wingate:2000bb,Lucini:2004my}. A more extensive discussion on the topic can be found in the \cite{Giataganas:2015ksa} and references therein.

\section{Confinement in Gauge/Gravity Duality}

Let us start with a generic background and place there a probe to check the presence of confinement. The probe is the static heavy meson Q\={Q} bound state where in the space-time it generates an orthogonal Wilson loop, with one edge being the spatial distance between the quarks $L$ and the other larger edge $T$ the time, satisfying $T\gg L$.

The background is homogeneous with the metric elements depending on the
holographic direction. It has the form
\be\label{metric1}
ds^2=-g_{00}\prt{u}dx_0^2+g_{11}\prt{u}dx_1^2+g_{22}\prt{u}dx_2^2+ g_{33}\prt{u}dx_3^2 +g_{uu}\prt{u}du^2 +\ldots
\ee
where the dots refer to metric elements of the rest of the space and include the internal space. This part of the space does not play any role in our computations since the probes there are localized consistently. This is due to the fact that the equations of motion of the string world-sheet we consider, are decoupled between the internal space and the rest of the space and as a result the string can be consistently always localized in the internal space. Notice that our metric \eq{metric1} accommodates as special cases the anisotropic homogeneous spaces.

The string world-sheet corresponding to the orthogonal Wilson loop is parametrized as
\be\label{ansatz111}
x_0=\t~,\qquad x_p=\sigma,\qquad \mbox{and}\qquad u=u(\s)~,
\ee
where we have chosen the static gauge and have allowed the string to enter the bulk.  The $x_p$, is the space direction along which the pair is aligned, it has metric element $g_{pp}$ and can be chosen as we like without any loss of generality. Notice that the translational invariance of the string along the time direction is essential for the
computation. The shape of such string can be found by solving ordinary differential equations and not partial differential equations. The approximation is valid since we have taken the time edge to be much larger than the spatial edge.

From the Nambu-Goto action we can derive the length of the Wilson loop in the boundary in terms of $u_0$, the deeper distance of the world-sheet in the bulk.  The length of the two endpoints of the string is given by
\be\label{staticL1}
L=2\int_{u_{0}}^{\infty}  du \sqrt{\frac{- g_{uu} c^2}{(g_{00}g_{pp}+ c^2)g_{pp}}}~,
\ee
where $c$ is a constant satisfying
\be
g_{00}\prt{u_0}g_{pp}\prt{u_0}=- c^2~.
\ee
The energy of the string after subtracting the mass of the two free quarks is
\bea\nn
&&2\pi\a' E= \sqrt{g_{00}\prt{u_0} g_{pp}\prt{u_0}} L+2\left[  \int_{u_{0}}^{\infty} du \sqrt{- g_{uu}g_{00}}\prt{\sqrt{1+\frac{c^2}{g_{pp}g_{00}}}-1}- \int_{u_{k}}^{u_0} du \sqrt{-g_{00}g_{uu}}\right]~,\\
&&~\label{energia1}
\eea
where $u_k$ is the horizon or an other cut-off of the metric. The first term of the expression \eq{energia1} may act as an effective string tension $T_{eff}$. In fact if this factor is non-zero, has a minimum at the bottom of the geometry and is monotonic with the holographic coordinate $u$, then the other integrals in the energy are exponential corrections to the linear term \cite{Sonnenschein:1999if}. This is the case when our gravity background is dual to a confining theory.

Confinement may be achieved using the bottom-up hard-wall models where a cut-off is introduced at some finite radius $u_{hwall}$ as in \cite{Polchinski:2001tt,Erlich:2005qh,deTeramond:2005su} \footnote{Later bottom-up models include \cite{Gursoy:2010fj}, while other studying in some depth the string tension include
\cite{Andreev:2006ct,BoschiFilho:2006pe,Andreev:2006nw}.}. These models are not solutions of the supergravity equations and therefore the correspondence to the dual gauge theory may be problematic. However, the bottom-up models, have proven useful and considering the fact that the computations are more tractable than in the more involved top-down models and the results can be obtained analytically, they can teach us interesting things for the gauge/gravity correspondence.

Another way to introduce confinement is in a top-down way, where the backgrounds used are coming as solutions of the supergravity equations of motion. The cut-off in these geometries is introduced by an extra dimension which shrinks to zero size at a holographic radial distance $u_k$ from the boundary. This geometry is sometimes called a cigar-type geometry because of its shape. The most commonly used confining background is the D4 Witten background \cite{Witten:1998zw}
\bea\label{ss1w}
&&ds^2=\left(\frac{u}{R}\right)^{3/2}
\prt{\eta_{\mu\nu}dx^\mu dx^\nu+f(u)dx_4^{2}}
+\prt{\frac{R}{u}}^{3/2}
\prt{\frac{du^2}{f(u)}+u^2 d\Omega_4^2}~,
\\\nn
&&e^\phi= g_s \left(\frac{u}{R}\right)^{3/4},
~F_4=dC_3=\frac{2\pi N_c}{V_4}~\epsilon_4 \ ,
~f(u)=1-\frac{u_k^3}{u^3} \ ,~R^3=\pi g_s N_c l_s^3 ~,
\eea
where $\mu=0,..,3$. $d\Omega_4^2$, $\epsilon_4$ and $V_4=8\pi^2/3$ are the line element, the volume form and the
volume of a unit $S^4$, respectively. The string length is $l_s = \sqrt{\a'}$ and the couplings satisfy the following relations:
\bea\nn
&& g_5^2=(2\pi)^2 g_s l_s~,\quad g_4^2=
\frac{g_5^2}{2\pi \rho}=3\sqrt{\pi}\left(\frac{g_s u_k}{N_c l_s}\right)^{1/2},\quad \lambda_5=g_5^2 N_c~.
\eea
There exist other top-down confining backgrounds with more involved geometries. In this short review we will not mention them, while some of them may be found in the review \cite{Aharony:2002up}.

Our theories do not have matter in the fundamental representation and therefore, several properties of confinement can not be studied with
the models presented so far. To add these fundamental degrees of freedom we need to introduce in our background higher dimensional flavor branes that touch the horizon of the original supergravity background \cite{Karch:2002sh,Erdmenger:2007cm}. Strings ending on the original branes and the flavor branes correspond to quarks in the fundamental representation, while open string ending between only the flavor branes correspond to new lighter mesonic degrees of freedom.

Such a confining example is  the Witten-Sakai-Sugimoto model \cite{Sakai:2004cn}, which consists of the near horizon limit of the D4-branes mentioned above and the additional flavor D8 branes. In this model the set of D8-branes touch the boundary at two points and can be thought to represent a $U\prt{N_f}\times U \prt{N_f}$ global symmetry, which is the chiral symmetry of fermions. The action for the D8-brane in the background \eq{ss1w} specifies the shape of the branes that have to be connected smoothly close to the bottom of the cigar-type geometry. This reduces the group to a single $U \prt{N_f}$ global symmetry which corresponds to the chiral symmetry breaking.

Using the flavor models described above we can study the meson spectrum, and any other observables where the fundamental degrees of freedom are affected by the gluonic ones. This is the probe limit of the D7 flavor branes. In other words, the number $N_f$ of the flavor branes is much smaller than the number of colors $N$, and the heavy branes do not backreact to the initial background. This corresponds to the quenched approximation in the dual field theory, where the fermionic determinant is neglected during the computation of the relevant observables.

In this approximation the screening of the static potential and the effective string breaking can not be computed since the virtual quarks are not present. To observe these phenomena in the context of AdS/CFT one needs to consider backreacting flavor branes on the geometry, which is dual to the unquenched limit in field theory. This can be done in the Veneziano limit
\be
N_c\rightarrow \infty~,\quad N_f\rightarrow \infty~,\quad \frac{N_f}{N_c}\rightarrow \mbox{fixed},\quad \lambda=g_{YM}^2 N_c \rightarrow fixed~.
\ee
The static potential computed in holographic models that exhibit realistic behavior regarding the fundamental quarks, is expected to fully qualitatively reproduce the phenomena associated to screening and the QCD string breaking. Due to certain difficulties to obtain such holographic models, the problem is still considered as open.

In this review we focus mostly on the properties of confinement using holography in the quenched approximation. Nice reviews on several topics of the large $N$ gauge theories from the lattice point of view include \cite{Teper:2009uf,Lucini:2012gg,Panero:2012qx}

\section{Properties of Confinement in Holography}

\subsection{The String tension}

As we have already described the string tension of a mesonic bound state in holography is specified by computing the expectation value of the orthogonal Wilson loop. This has been done in several cases in the zero temperature confining models applying the formalism used to obtain the equation \eq{energia1}.

However, one may try to question the dependence of the string tension on the temperature while in the confining phase. Before studying the problem in holography, let us first review quickly the main details of the string model computation. We require all the bosonic fields to be periodic with the $\beta=T^{-1}$, and the boundary of the world-sheet to be
\be
X\prt{0,\tau}=X\prt{L,\tau}=0~,\quad X\prt{\s,0}=X\prt{\s,\beta}~,
\ee
with $L\gg \beta$. In the limit of low temperature  the expansion in lowest order of $T^2$ gives
\be\label{sigma11}
\sigma\prt{T}\simeq \sigma_0-\frac{\pi\prt{d-2}}{6} T^2+\ldots~.
\ee
Notice the decrease of the string tension with the temperature, an effect which is naturally expected. Moreover, the first temperature dependent term is related to the universal L\"{u}sher term in the linear potential at zero temperature. In principle this can be worked out in the context of Wilson loops with appropriate boundary conditions, in backgrounds of AdS/CFT with confinement and to our knowledge this problem is not fully solved so far.

On the other hand, there are other ways to compute the string tension dependence on the temperature in the gauge/gravity duality. One may compute the higher order temperature dependent corrections of the supergravity solution itself, while in confined phase. This is a difficult task, and we are not aware of any such top-down solution.

A different but equally interesting approach is to bring the probe string in the thermal confining background, to thermal equilibrium with the heat bath. This has been achieved in \cite{Giataganas:2014mla} in the top-down approach that computes the string tension of the heavy quark bound state using the blackfold methodology
\cite{Emparan:2009cs,Emparan:2009at}. For
our purpose we use  blackfolds to capture the thermal
excitations of the string world-sheet corresponding to a Wilson loop
in a heat bath. The probe black string is placed in
a finite temperature background in thermal equilibrium with it. This method has been also used  for  Wilson loops in  finite temperature
$AdS$ and non-relativistic spaces \cite{Grignani:2012iw,Armas:2014nea}.

The expectation value of the Wilson loop is computed by using the free energy of
the probe, which in the zero temperature limit reproduces the result of the traditional Nambu-Goto action approach \cite{wlf}.  The bound state considered consists of $k$ separate, fundamental strings with $1\ll k\ll N$. This is not to be confused with $k$-strings which correspond to Wilson loops in higher symmetric and antisymmetric representations and in the dual gravity theory are represented as $D3$ and $D5$ branes respectively with supporting electric fluxes \cite{wld5}.

The methodology developed in \cite{Giataganas:2014mla} was applied in the thermal soliton $AdS_3$ background in the low temperature phase
corresponds to a confining theory.  The background reads
\bea\label{metrica11}
&&ds^2=\frac{u^2}{R^2}\prt{-dt^2+dx_1^2+dx_2^2+f\prt{u}d\phi^2}+ \frac{R^2}{u^2 f\prt{u}}du^2~,~\\
&&f(u)=1-\frac{u_k^4}{u^4}~.~\label{eq:f11}
\eea
The tip of the cigar like topology is at $u=u_k$ and to avoid the singularity at the tip, certain periodicity conditions are imposed: $\phi \sim \phi+ 2\pi \rho~,~ \rho:=
R^{2}/u_{k}~.$
The Euclidean time $t$ has a periodicity associated to the
inverse temperature $1/T:=2 \pi u_h$. In the confining phase the time circle never shrinks to zero, while the
circle in the $\phi$ direction shrinks to zero at the tip.


The static potential has the linear term indicating confinement and the higher order corrections depending exponentially on the size $-L$ of the bound state. In this limit the leading contributions come from the region close to the turning point, where the string almost totally lies when $u_0\simeq u_k$ \cite{Sonnenschein:1999if,Giataganas:2011nz,Giataganas:2011uy,Nunez:2009da}.

We parametrize the $F$-string probe with the radial gauge $t=\tau,$ $u=\sigma$ and $x=x\prt{\sigma}$. In a generic background the free energy and the size L of the bound state are \footnote{Here we do not present the computational technicalities of the blackfold to ease the presentation and we focus mostly on the results and properties of the confinement. The details can be found in the original work \cite{Giataganas:2014mla}.}
\bea \label{fenergy}
&&\cF=A \left(\frac{3}{2\pi T}\right)^6\int_{\s_b}^{\s_0} d\s
\sqrt{f_u(u)}\sqrt{1+ f_{1u}(u)
  x'^2}|g_{00}|^3\frac{1+6\sinh^2\a}{\cosh^6\a}~,\\
\label{lll1}
&&\frac{L}{2}=\int_{\s_b}^{\s_0}d\s\left(\frac{ f_{1u}^2(\s) G^2(\s)}{
    f_{1u}(\s_0)G^2(\s_0)}-f_{1u}(\s)\right)^{-1/2}~.
\eea
where
\bea
\label{deffs}
\quad f_u(u):=|g_{00}(u)| g_{uu}(u)~,\quad
f_{1u}(u):=\frac{g_{11}(u)}{g_{uu}(u)}~,\quad G(\s):=\sqrt{f_u}~|g_{00}|^3 \frac{1+6\sinh^2\a}{\cosh^6\a}~.
\eea
The parameter $\a$ is the dimensionless charge parameter, which turn out to be equal to
\be
\a \simeq\frac{1}{4}\log\left(\frac{16 B(\s)}{\kappa}\right)-\frac{3\sqrt{\kappa}}{8\sqrt{B(\sigma)} }+\cO \prt{\frac{\kappa}{B(\sigma)}}~,
\ee
with
\be
\label{betadef}
\kappa=\frac{2^7}{3^6}\frac{k \sqrt{\lambda}}{N^2}~,\qquad B(\sigma)=\frac{|g_{00}|^3}{\pi^6 T^6 R^6}~.
\ee
$R$ is the radius of the space and $T$ is the temperature of the heat bath. The number $k$ corresponds to the $k$ separate
Q\={Q} bound states, that do not interact at zero temperature.

We consider the approximation of large $L$, and by subtracting the infinities \cite{Chu:2008xg,Drukker:1999zq} we obtain the regularized energy expression to find  the string tension
\be\label{ell21}
\cF_{linear}=\frac{\sqrt{\lambda}k}{2\pi R^2}\left(\prt{\frac{u_k}{R}}^{2} -\sqrt{\kappa}\frac{\pi^3 R^{4} T^3}{3 u_k} \right)L~,
\ee
where $\s_0\propto u_k^2/R^2$, the string tension at zero temperature. This is a naturally expected result: the string tension of the QCD string between the heavy quark bound state is decreased as the temperature is increased.
The first temperature dependent term in the tension is of $T^3$ order.
This differs to the result of the heavy meson string tension dependence with temperature presented in \eq{sigma11}.
Our bound state may be interpreted as a state of non-interacting quarks pairs at zero temperature. By increasing the temperature we find the energy of the $k$ separate strings in thermal equilibrium with the background. However, we see that the thermal terms in the
energy depend on the number of strings, in a way that can not be factorized as a common factor in the expression. While bringing the separate strings into thermal equilibrium with the
background, the strings appear to interact although this interaction may be indirect. The limit to the single meson state can not be taken, since the blackfold approach is not valid anymore, so a direct quantitative comparison can not be done.

Therefore the blackfold obtained result, need not to necessarily match with the $T^n$ scaling of the string tension for
the single quark pair found using effective string models given the differences between the system of two bound states, and the physics that they probe, but the results might be  complementary.

\subsection{The Width of The Flux Tube}
\subsubsection{Connected Minimal Surface}

The width of the flux tube can be measured by the connected minimal surface of the Q\={Q} large Wilson Loop $C$ and the probe small Wilson loop $c$. For our purposes the orthogonal Wilson loop can be taken as a cyclic to ease the computations, without affecting the results in the limit we work. The two cyclic Wilson loops of radii $R_2\gg R_1$ need to be placed on the boundary of the confining theory, say on the spatial plane $x_1 x_2$ with the same center and along the direction $x_3$ are separated by a distance $L$.

\begin{figure*}[!ht]
\begin{minipage}[ht]{0.5\textwidth}
\begin{flushleft}
\centerline{\includegraphics[width=80mm]{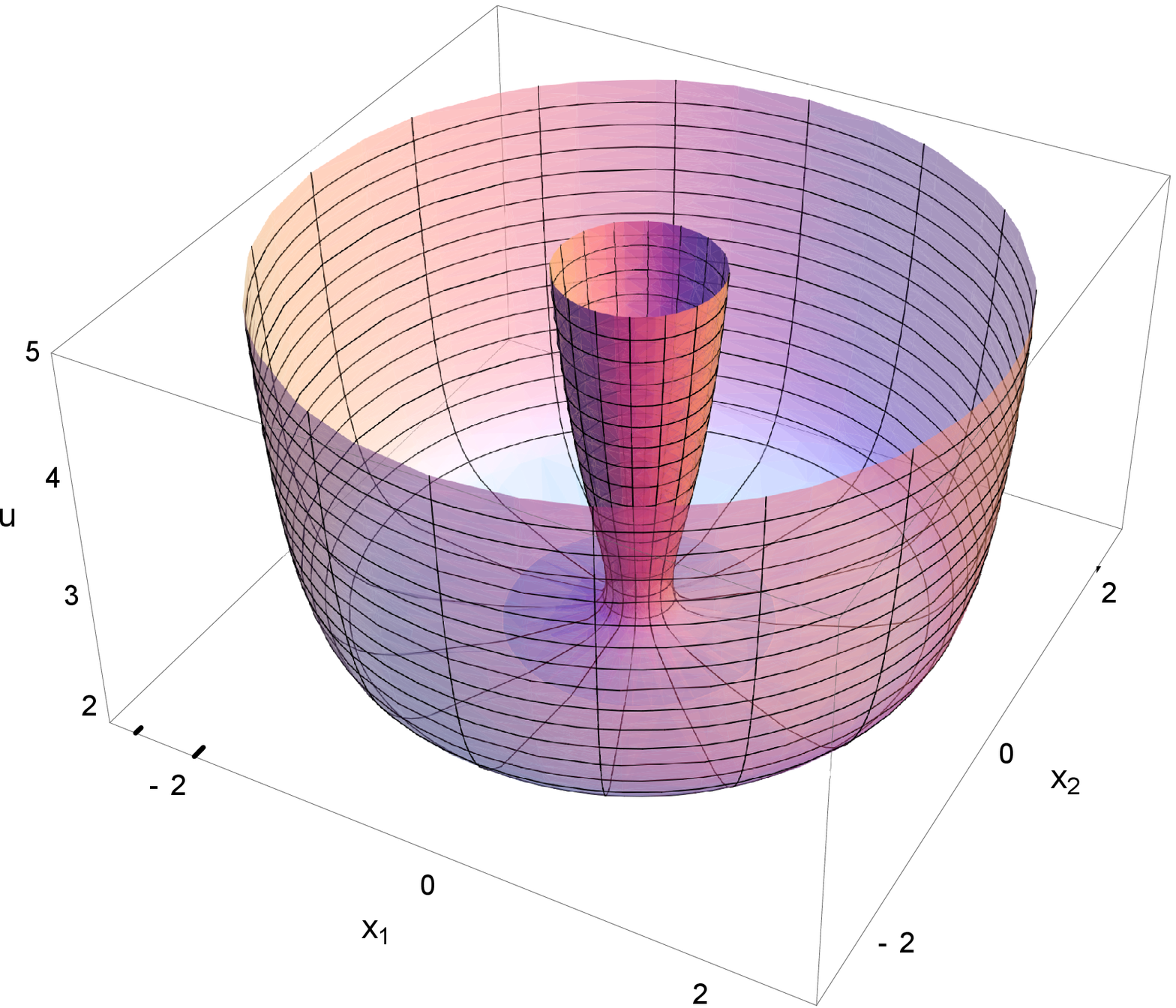}}
\caption{\small{The minimal surface in the $\prt{x_1=r \cos \theta, x_2=r \sin\theta,u}$ space. We observe the two cyclic Wilson loops with the common center on the boundary. As the surface attempts to minimize, it enters in the bulk.  The change in the radius of the surface,  happens almost totally around the tip of the geometry $u_k$.
\vspace{-0.1cm}}}\label{fig1}
\end{flushleft}
\end{minipage}
\hspace{0.4cm}
\begin{minipage}[ht]{0.5\textwidth}
\begin{flushleft}
\centerline{\includegraphics[width=80mm]{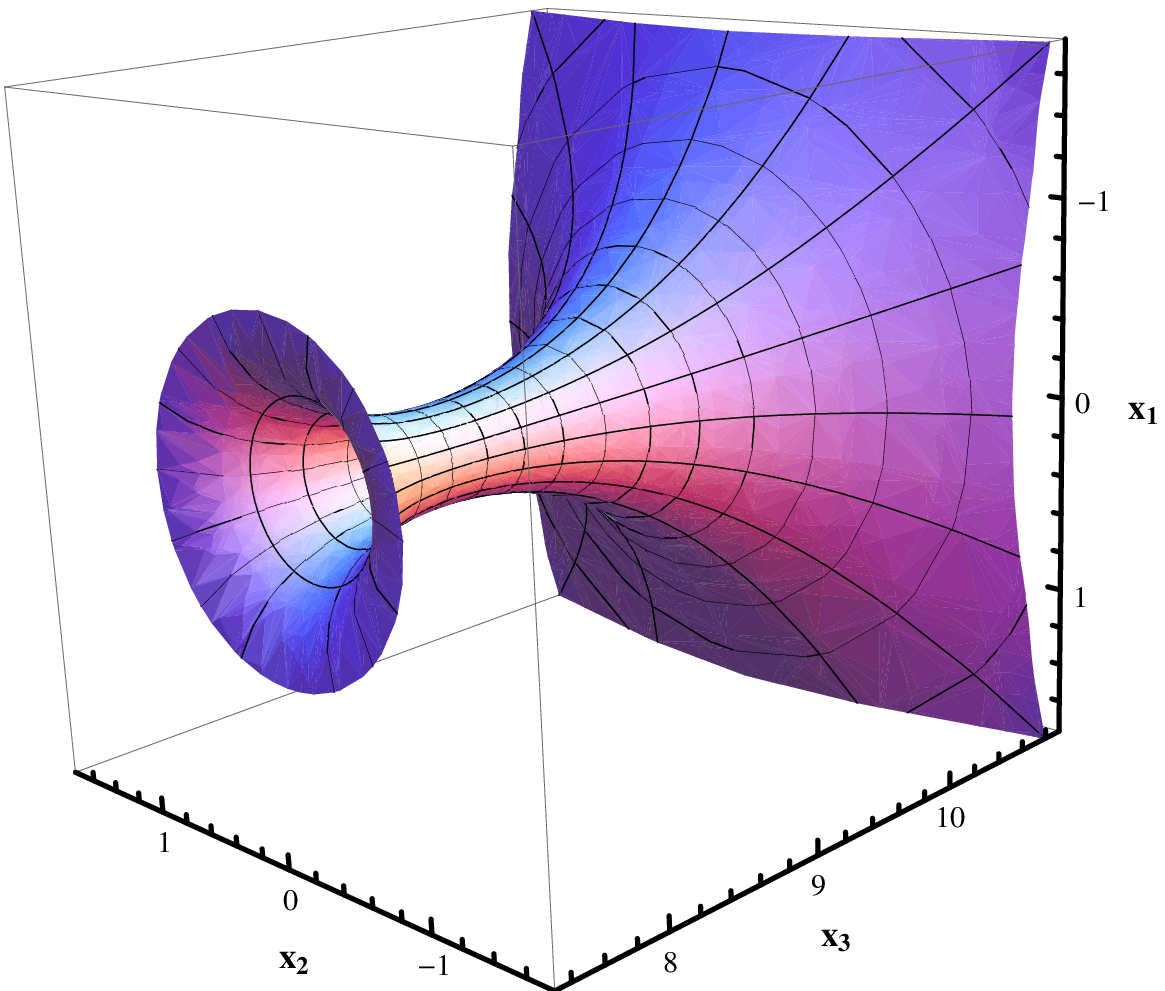}}
\caption{\small{The minimal surface in the $\prt{x_1,x_2,x_3}$ space. We observe that is a rapid change of the size of the tube as it departs from its boundaries.  In the particular plot, the spatial $x_3$ extension of the tube is five times enhanced for presentation purposes.
\vspace{0.5cm}}}\label{fig2}
\end{flushleft}
\end{minipage}
\end{figure*}

In the following we describe very briefly some of the generic results of \cite{Giataganas:2015yaa} \footnote{In the bibliography there are few holographic studies on the flux tube, or  configurations that related to it, for example \cite{widtha11,Olesen:2000ji}.}. By making a coordinate transformation on the metric \eq{metric1} we work with the convenient form
\be\label{metriccca1}
ds^2=-g_{00}\prt{u}dx_0^2+g_{11}\prt{u} \prt{dr^2+r^2 d\theta^2 +dx_3^2}+g_{uu}\prt{u}du^2+dX^5~.
\ee
The string world-sheet we have described is parametrized by
\be\label{WLprmt}
\theta=\tau~,\quad x_3=\sigma~,\quad r=r\prt{\sigma}~,\quad u=u\prt{\sigma}~.
\ee
The function $r\prt{\sigma}$ specifies the diameter of the formed
tube between the Wilson loops, while $u=u\prt{\sigma}$ gives the profile of the surface in the bulk. The Nambu-Goto action
\be
S=\frac{1}{2\pi \a'}\int d\s d\t g_{11}r\sqrt{1+r'^2+\frac{g_{uu}}{g_{11}}u'^2}
\ee
gives a system of differential equations
\bea
&&1+r'^2+\frac{g_{uu}}{g_{11}}u'^2-\frac{g_{11}^2 r^2}{c^2}=0~,\\
&&r''-\frac{g_{11}^2 r}{c^2}=0~,\\
&&u''\frac{g_{uu}}{g_{11}}+u'^2\frac{1}{2} \partial_u\prt{\frac{g_{uu}}{g_{11}}}-\frac{r^2 g_{11} \del_u g_{11}}{c^2}=0~,
\eea
where $c$ is a constant and the first equation can be thought as a constraint. The system has an analytic solution only in the case of the $AdS$ backgrounds since total derivatives are formed \cite{Olesen:2000ji,Giataganas:2015yaa}. In all the other cases we need to employ numerics or to consider sensible approximations in order to get analytic solutions. The two-dimensional minimal surface projections of the exact numerical solutions in the D4 Witten model \cite{Witten:1998zw} have been found in \cite{Giataganas:2015yaa} and are presented here in Figures \ref{fig1} and \ref{fig2}.

To obtain analytic solutions we consider particular limits. By considering large flux tubes with $R_2\gg R_1$ it turns out that
\be
L= \frac{c}{g_{11}\prt{u_k}} \log \frac{R_2}{R_1}~,
\ee
where $u_k$ is the tip of the cigar type geometry along the holographic direction and the turning point $u_0$ of the surface is very close to the tip of the geometry, $u_0 \simeq u_k$ . In this case we know that the shape of the different projections of the string world-sheet takes a rectangular $\Pi$ shape departing from the boundary and entering to the bulk steeply. The action can be found to approach \cite{Giataganas:2015yaa}
\be
S\simeq
\s\prt{\frac{L^2}{\log\frac{R_2}{R_1}}+R_2^2-R_1^2}\, .
\ee
Using the equation \eq{wdefen11} and \eq{wdefen22} we show that all confining holographic models have the property of the logarithmic broadening
\be
w^2\propto\log\frac{R_2}{R_1}~,
\ee
where the position of the small loop along the $x_3$ direction does not play any role as expected.

\subsubsection{String Fluctuations}

An alternative way to compute the width of the flux tube is to consider time-dependent transverse fluctuations of the string connecting the static heavy quarks. The width is given by the expectation values of the square transverse deviations of the string, e.g. as in \cite{Gliozzi:2010zv,Gliozzi:2010zt}
\bea
w^2(x) &=& \langle \delta x(\sigma,\tau)^2 \rangle \equiv \lim_{\e\to 0}\langle \delta x(\sigma,\tau) \delta x(\sigma'=\sigma+\e, \tau'=\tau+\e) \rangle\nonumber\\
&=&  \lim_{\e\to 0}G(\sigma,\tau;\sigma'=\sigma+\e,\tau'=\tau+\e))\, \label{flucdef1}
\eea
where $G$ is the appropriate Green function.

The corresponding computation is involved even in the flat space. In the limit of large compactified Euclidean time $\beta$ compared to the distance of the bound state $L$, $\beta\gg L$, the width is given by \cite{Gliozzi:2010zv,Gliozzi:2010zt,Gliozzi:2010jh}
\be\label{widthwa1}
w^2\prt{L/2}=\frac{1}{2\pi\sigma}\log{\frac{L}{L_0}}+\frac{1}{\pi\sigma}\log \eta\prt{ 2\tau}\, ,
\ee
where $\eta$ is the Dedekind function.

In a curved space, the Nambu-Goto action after the perturbation
of the Wilson loop
\be
t=\t~,\quad x_1=x_1\prt{\sigma}+\delta x_1 \prt{\sigma}~, \quad x_2=\delta x_2\prt{\sigma}~, \quad u=\sigma~,\label{pert1}
\ee
becomes
\be
S_2=\frac{1}{2}\int d\sigma \prt{\frac{g_{00}^2 g_{uu} g_{11}}{ F^{3/2}}\delta x_1'^2+\frac{g_{00} g_{22}}{\sqrt{F}}\delta x_2'^2-\frac{g_{11} g_{uu}}{\sqrt{F}}\delta \dot x_1^2- \frac{g_{22}  \sqrt{F}}{g_{00}}\delta \dot  x_2^2}~ ,
\ee
where
\be
F:=\frac{g_{00}^2g_{uu}g_{11}}{g_{00}g_{11}-g_{00_0}g_{11_0}}
\ee
and $g_{00_0}:=g_{00}\prt{\sigma_0}$, $g_{11_0}:=g_{11}\prt{\sigma_0}$. The turning point of the string is $\sigma_0$.
The general equations of motion an be found
\bea
&&\frac{d}{d\sigma}\prt{\frac{g_{00}^2 g_{uu} g_{11}}{ F^{3/2}}\delta x_1'} -\frac{g_{11} g_{uu}}{\sqrt{F}} \delta \ddot x_1=0\, , \\\label{e212}
&&\frac{d}{d\sigma}\prt{\frac{g_{00} g_{22}}{\sqrt{F}}\delta x_2'}-\frac{g_{22} \sqrt{F}}{g_{00}} \delta\ddot x_2=0~,
\eea
valid for any diagonal homogeneous background of the form \eq{metric1}. The analytic form of the Green function of the differential equation \eq{e212} of the transverse fluctuations with appropriate boundary conditions, in a confining background is a difficult problem to be solved. A discussion on the subject can be found in \cite{Giataganas:2015yaa}.

\subsection{The $k$-strings}

The single string corresponds to the Wilson loop in the fundamental representation \cite{wlf}, while for Wilson loops in higher symmetric and antisymmetric representations are appropriate the D3 branes and the D5 branes with electric flux  \cite{wld5}. There are several holographic studies on the $k$-string tension, the L\"{u}sher term and other observables, for example the references in \cite{kstrings1a1}. Moreover, in \cite{univwl} it has been shown that the energy of higher representation Wilson loops in very special backgrounds with trivial dilaton is expressed in terms of the energy of the loop in the fundamental representation.

It turns out that this observation is true for a much wider class of backgrounds. In \cite{Giataganas:2015ksa} it has been found that the observables of $k$-strings can be expressed in terms of observables of single strings when a condition of the background is satisfied. This means, for example, that the string tension dependence of the $k$-string on $\s_1$ in these theories can be read directly from the final result in the analysis that follows.

Let us consider the $d+1$ dimensional homogeneous space with the metric
\bea
ds^2 = g_{00}\prt{u} dx_0^2+g_{ii}\prt{u}dx_i^2  +g_{uu}\prt{u} du^2+ g_{\cX}\prt{u}\left(d\theta^2+s_\theta^2 d\cX_q^2\right)\, ,
 \label{met1a1}
\eea
where $i=1,\ldots,d$ and the internal space consists of the angle $\theta$ and the $q$ dimensional manifold $\cX$. The background has additional non trivial dilaton and appropriate forms. Let us work with $q=4$ and $d=3$ dimensional spaces without loss of generality. In this theory we place a $D5$-brane describing the heavy bound state. Its extension in the space-time should be as the one of the fundamental Wilson loop and it should also wrap the $\cX$ manifold. It is parametrized by
\be
x_0=\tau~,\quad x_1=\sigma~, \quad u=u(\sigma)~,\quad\theta=\theta(\sigma)\quad \mbox{and wrapping  \cX${}_4$}~,
\ee
The action for the D5-brane is
\be\label{dbia1}
S=T_{D_5}\int d\tau d^5\sigma e^{-\phi}\sqrt{g+ 2  \pi \a' F_{\m\n}}-i g_{st} T_{D_5}\int 2 \pi \a'F_{\mu\nu}\wedge C_4~,
\ee
where
\be
\lambda=g_s N~, \quad T_{D_5}=\frac{N\sqrt{\lambda}}{8 \pi^4}~,\quad
C_4:=-\frac{D\prt{\theta}}{g_{st}}vol_X~
\ee
and $vol_X$ is the volume form of the $\cX$ space.  The electric flux is rescaled $F_{\tau\sigma}=i F/2\pi \alpha'$
and the equations of motion of the brane come with and additional charge quantization condition
\be
\frac{\delta S}{\delta F}=-\frac{k}{2\pi \alpha'}~.
\ee
In order to express the D5-brane action in terms of the Nambu-Goto action, we need to find a constant $\theta$ solution to the equations of motion
\bea\label{thetaeom}
&&4 s_\theta^3 c_\theta h \sqrt{G_s-F^2}-F \partial_\theta D  =\partial_1\left(\frac{s_\theta^4 g_{00}g_{\theta\theta}\theta'}{\sqrt{G_s-F^2}}\right)~,\\
&&\sqrt{G_s-F^2}=\frac{s^4_\theta h F}{- D+\tilde{k}}~,\label{Feom}
\eea
where $h\prt{u}:=e^{-\phi}  g_{\cX}^2$, $G_s$ is the induced metric of the fundamental string and $\tilde{k}$ is a rescaled number of the Q\={Q} pairs:
\be
G_s=g_{00}\prt{g_{11}+g_{uu}u'^2 +g_{\theta\theta}\theta'^2}~\quad\mbox{and}\quad \tilde{k}:=\frac{4\pi^3 k}{N Vol_X }~.\label{eomf}
\ee
It turns out that the equation of motion for $\theta$ becomes \cite{Giataganas:2015ksa}
\be\label{thetaequation}
4 h^2 c_\theta s_\theta^7+ D \partial_\theta D- \tilde{k} \partial_\theta D =0~,
\ee
which has non-trivial solutions for $\theta=\theta_0$  when the function $h\prt{u}$ satisfies
\be\label{conditiond1}
e^{-\phi\prt{u}} g_\cX\prt{u}^2 =c~,
\ee
where the metric cancels the dilaton contribution. This condition is satisfied by several top-down backgrounds, even anisotropic ones \cite{Giataganas:2015ksa}. The condition \eq{conditiond1} is part of the expression $e^{-\phi} \sqrt{g}$, with $g$ being the induced metric related to the wrapping of the internal space. This is known to be conserved under T-dualities, and therefore it is natural to expect that should be independent of the holographic coordinate $u$, in order to be able to map at the level of equations of motion the T-dualized action of \eq{dbia1} to the Nambu-Goto action.

The system can be solved analytically for constant $\theta$ provided that the condition \eq{conditiond1} is satisfied.  The electric flux is found to be
\be\label{electric}
F= Z\prt{\theta}\sqrt{G_s}~,
\ee
where
\be
Z\prt{\theta}:=\frac{\sqrt{G_s}}{\sqrt{1-\frac{s_{\theta} \partial_{\theta} D}{4  c_{\theta}}}}~.
\ee
The action of the string needs to be regularized due to the present divergences \cite{Chu:2008xg,Drukker:1999zq,wldd} by adding the following terms to the action
\be\label{boundarya1}
S_{D_5~b}=-\int_\partial d\sigma x^\mu\frac{\delta S_{D_5}}{\delta\partial_\sigma x^\mu}+\frac{\sqrt{\lambda}}{2 \pi}\int d\tau d\sigma k F~.
\ee
The first term is the string action infinity subtraction term, where $x^\mu$ is taken as the radial direction. By using the equations of motion and subtracting the infinities we end up with a compact result
\be\label{finalf11}
S_{D5}
= \frac{N Vol_\cX}{4 \pi^3} \frac{s_\theta^4}{ \sqrt{1-Z\prt{\theta}^2}} S_{NG,normalized}~.
\ee
Notice that the angle $\theta$ is given in terms of the number $k$ by solving the equation \eq{thetaequation}, when the \eq{conditiond1} is satisfied. Therefore, the energy of the brane configuration is proportional to the energy of the fundamental string configuration in the same space, with a constant of proportionality depending on $k$.

Theories that satisfy the condition \eq{conditiond1}, have $k$-strings observables like the dragging of the $k$-string in the quark-gluon plasma that can be written as  proportional to the observables of the fundamental strings. The treatment of the $k$-string observables can be done in full generality for the different backgrounds like the fundamental string observables following the analysis of \cite{Giataganas:2012zy,Giataganas:2013lga}. Moreover, several other features of the bound states should be possible to be expressed in terms of the fundamental ones. For example, the width of the $k$-string should be proportional to the fundamental string, as follows directly from the \eq{finalf11}.

\section{Concluding Remarks}

In this short review we have discussed properties of confinement and ways to probe it. Few new ideas and open problems have been also presented. The probes that have been discussed are the Wilson loop, the Polyakov loop and the t' Hooft loop. All three objects have unique features that signal the confined phase. Moreover we have mentioned the $k$-string bound states, involving several quarks and anti-quarks in an $SU\prt{N}$ theory. These states may be important in teaching us lessons for the dynamics of the confinement that the single meson states can not reveal. The debate for the form of the string tension of these states is quickly presented.

Looking at the confinement from the holographic point of view we have focused more on the Wilson loop operator. We have discussed a way to compute the dependence of the string tension on the temperature while the theory is still in the confined phase. There are certain difficulties to find supergravity top-down solutions in confined phase with temperature dependence in their metric. For this reason we use the blackfold approach to capture the thermal excitations of the string world-sheet corresponding to a Wilson loop in a heat bath. The probe string is brought to a thermal equilibrium with the background, introducing in the relevant analysis the heat bath temperature. The results are qualitatively to the correct direction, showing decrease of the string tension in presence of the finite temperature. We have also discussed the reasons of the discrepancies with the effective string models.

Then we have presented the width of the chromo-electric flux tube formed between a heavy meson bound state. To measure the width, two Wilson loops need to be inserted in the theory. The original Q\={Q} Wilson loop $C$ and a test small Wilson loop $c$ that measure the flux though it. The configuration to study in the holography is a connected minimal surface which ends at these two loops. We have shown that the logarithmic broadening of the flux tube is a property of all the confining theories. This is due to the common properties that the connected minimal surface has in all the confining backgrounds. We argue that there is an alternative way to compute the width of the flux tube through the time dependent transverse fluctuations of the original Wilson loop, in a way similar to the effective string models. The computation is highly involved due to the complicated differential equations of the fluctuations in the curved holographic spaces. A further question to be answered using holography is on the temperature dependence of the width of the flux tube while in the confining phase.

Finally we have presented properties of the $k$-string bound states. We have studied the string tension of the multi-quark bound state, and have found when it can be written in terms of the string tension of the meson bound state. It turns out that when the background theory satisfies a condition we provide, the static potential of the $k$-string is proportional to the potential on the fundamental string. In fact this is true not only for the static potential and string tension but for other observables, for example the dragging of the $k$-quarks moving in the quark-gluon plasma. It is not unexpected since looking macroscopically at the flux tube of this composite bound state, it is natural to behave as the usual QCD string with a different tension. However, we expect that higher order corrections in finite temperature theories will modify the proportionality relation and the $k$-strings will  be related in a more complicated way to the fundamental string.\newline~\\

\textbf{Acknowledgements:} The research of D.G is implemented under the "ARISTEIA" action of the "operational programme education and lifelong learning" and is co-funded by the European Social Fund (ESF) and National Resources.

\end{document}